# GLOBUS: Global building renovation potential by 2070

Shufan Zhang [1], Minda Ma [2 *, 3], Nan Zhou [3], Jinyue Yan [4]

1. School of Management Science and Real Estate, Chongqing University, Chongqing, 400045, PR China
2. School of Architecture and Urban Planning, Chongqing University, Chongqing, 400045, PR China
3. Building Technology and Urban Systems Division, Energy Technologies Area, Lawrence Berkeley National Laboratory, Berkeley, CA 94720, United States
4. Department of Building Environment and Energy Engineering, The Hong Kong Polytechnic University, Kowloon, Hong Kong, PR China

- Lead contact: Dr. Minda Ma (maminda@lbl.gov)

Summary

Surpassing the two large emission sectors of transportation and industry, the building sector accounted for 34% and 37% of global energy consumption and carbon emissions in 2021, respectively. The building sector, the final piece to be addressed in the transition to net-zero carbon emissions, requires a comprehensive, multisectoral strategy for reducing emissions. Until now, the absence of data on global building floorspace has impeded the measurement of building carbon intensity (carbon emissions per floorspace) and the identification of ways to achieve carbon neutrality for buildings. For this study, we develop a global building stock model (GLOBUS) to fill that data gap. Our study's primary contribution lies in providing a dataset of global building stock turnover using scenarios that incorporate various levels of building renovation. By unifying the evaluation indicators, the dataset empowers building science researchers to perform comparative analyses based on floorspace. Specifically, the building stock dataset establishes a reference for measuring carbon emission intensity and decarbonization intensity of buildings within different countries. Further, we emphasize the sufficiency of existing buildings by incorporating building renovation into the model. Renovation can minimize the need to expand the building stock, thereby bolstering decarbonization of the building sector.

# Supplemental Information

## 1. SUPPLEMENTAL DATA

**Trends in global floorspace per capita ignoring building renovation**

Figure S1 illustrates the trends in per capita floorspace from 2000 to 2070 for residential and non-residential buildings in 14 economies based on this study's non-renovation (NR) scenario. We developed the trends primarily by relying on databases maintained by reputable and authoritative institutions: (e.g., International Energy Agency (IEA),[1-3] United States (U.S.) Energy Information Administration,[4] United Nations Environment Programme,[5] and Eurostat [6]) and on highly cited journal articles authored by experts native to the specific economies.[7-9] Detailed reference information for each economy is presented in Section 3, Scenario Setting and Assumptions.

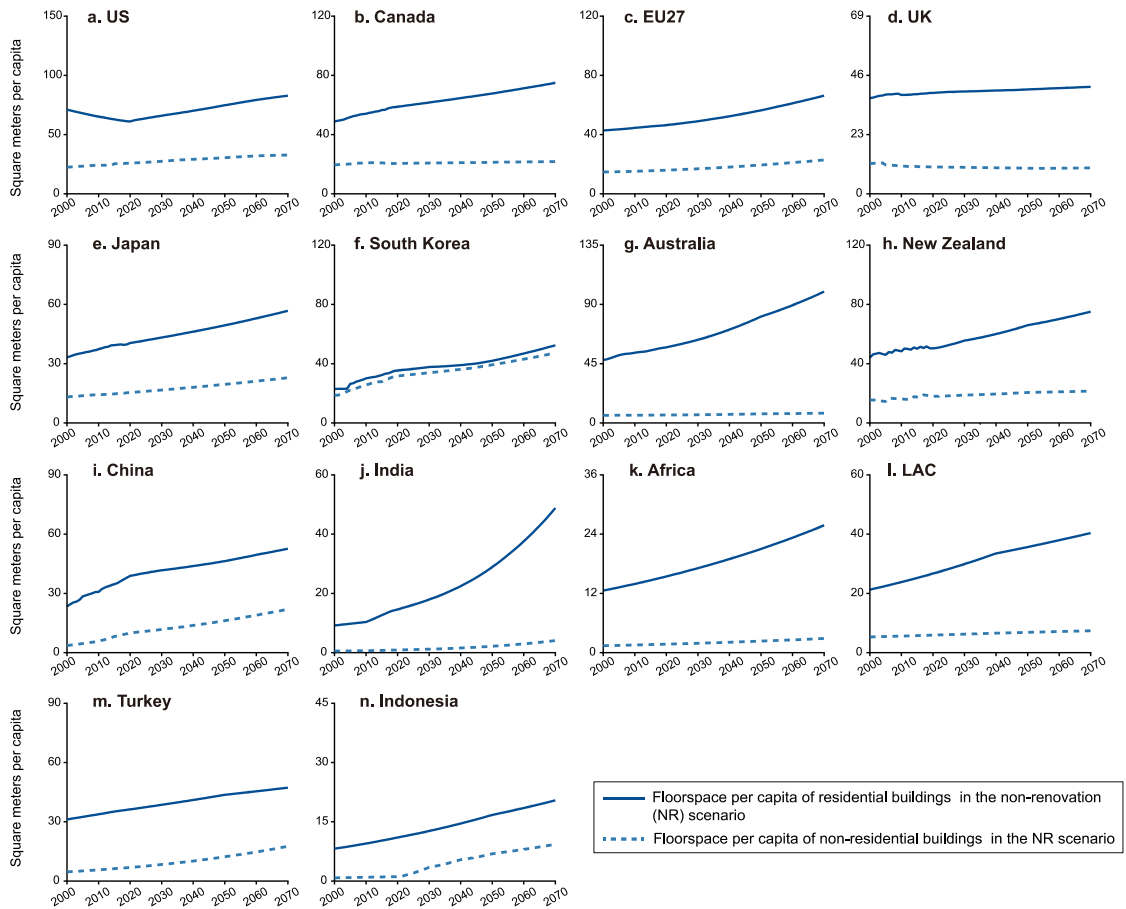

**Figure S1.** Trends in global floorspace per capita for residential and non-residential buildings from 2000 to 2070 based on the NR scenario.

Generally, future per capita floorspace in the 14 selected economies will exhibit an upward trend, as shown in Figure S1. In developed economies, the compound annual growth rates of per capita floorspace of residential buildings from 2000 to 2070 are predominantly within the range of 0.2% to 1.2%/yr, while the compound annual growth rates of the per capita floorspace of non-residential buildings generally fall between 0.2% and 1.3%/yr. Among developed countries, South Korea stands out for having the swiftest growth in per capita floorspace, attributable to a notable decline in population. The per capita floorspace of residential and non-residential buildings in South Korea will reach 52.4 and 47.2 square meters ($m^2$)*, respectively, by 2070. Australia and the US will be the top two developed countries in terms of per capita floorspace of residential buildings in 2070, reaching 99.7 and 82.8 $m^2$, respectively. New Zealand and Canada will also exceed 75 square meters per capita ($m^2$/person) by 2070. For non-residential buildings, US per capita floorspace will reach 32.7 $m^2$ by 2070, second only to South Korea. In Japan, the 27 countries of the European Union (EU27), Canada, and New Zealand, the per capita floorspace of non-residential buildings will exceed 21 $m^2$ by 2070.

For emerging economies, from 2000 to 2070 the compound annual growth rates of per capita floorspace of residential buildings range from approximately 0.6% to 2.4%/yr, while the compound annual growth rates of per capita floorspace of non-residential buildings range from 0.5% to 3.5%/yr. The per capita floorspace of residential buildings in India, the fastest-growing emerging economy, will reach 48.8 $m^2$ by 2070. By 2070, China will have the largest per capita floorspace of residential buildings of emerging economies, reaching 52.6 $m^2$. Additionally, by 2070 residential building stocks in Turkey and Latin America and the Caribbean (LAC) will have floorspaces of 47.3 and 40.4 $m^2$/person, respectively. With respect to per capita floorspace of non-residential buildings, Indonesia and India will reach impressive growth rates of 3.5% and 2.9%/yr, respectively. Even given those growth rates, by 2070 the per capita floorspace of non-residential buildings in Indonesia and India will remain relatively low, at 9.3 and 4.1 $m^2$, respectively, significantly lower than China's 22.0 $m^2$/person. Notably, after the 2020s the compound annual growth rates of per capita floorspace of residential and non-residential buildings in China will decrease to 0.6% and 1.6%, respectively.

---

* 1 square meter equals to 10.7639 square feet.

**Potential building stock changes considering renovation**

Building renovation has the potential to extend the service lifetimes of buildings and fulfill usage demands.[10,11] Accounting for future building renovations under the business-as-usual (BAU) and techno-economic potential (TEP) scenarios will produce lower increases in building stocks compared to results of the NR scenario (Figure S2). The disparities among the three scenarios are evident primarily in establishment of the building renovation rate and its corresponding impact on building lifetime.

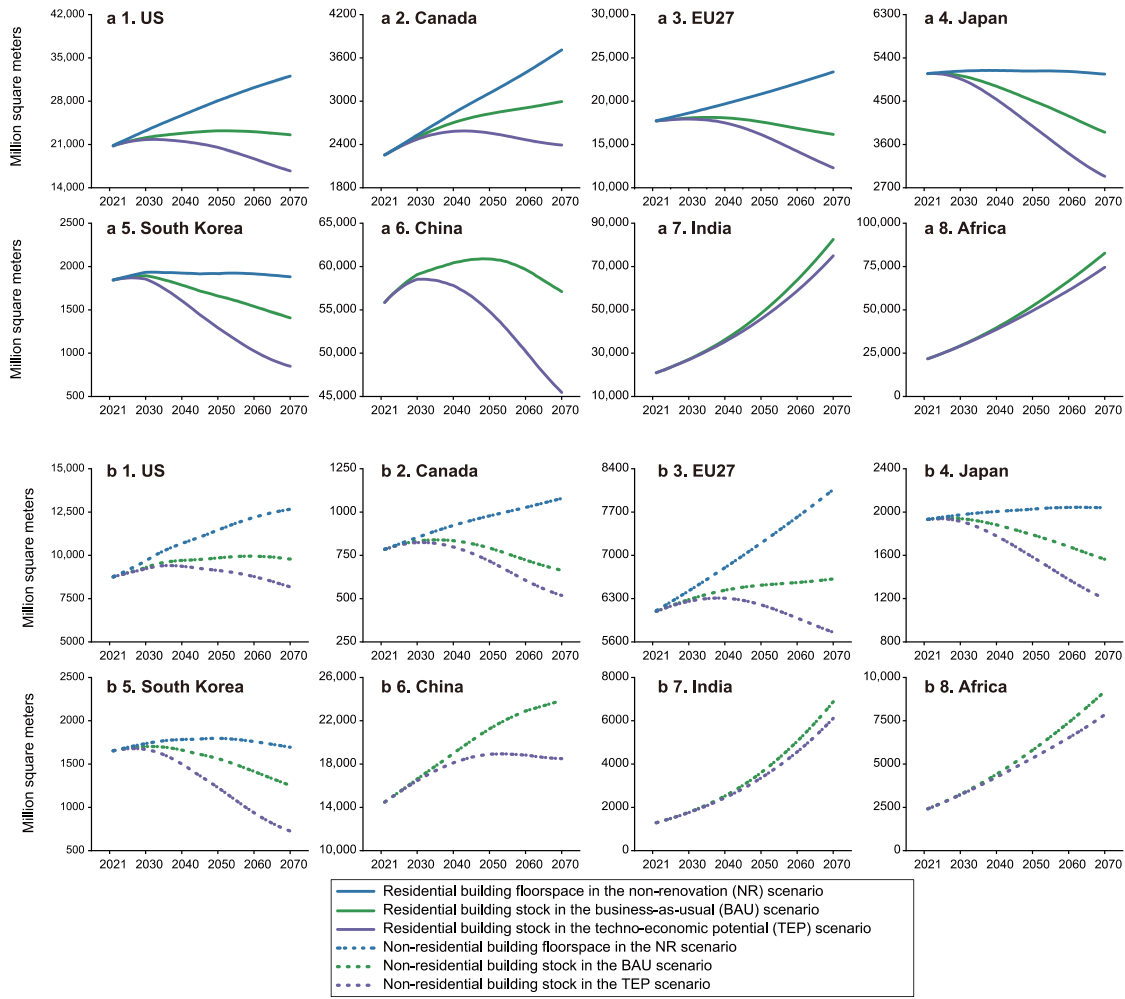

**Figure S2.** Trends in floorspace for (a) residential buildings and (b) non-residential buildings for eight economies based on the NR, BAU, and TEP scenarios, 2021–2070.

Among developed economies, the TEP scenario gives South Korea, the US, and the EU27 the greatest potential to substantially reduce residential building stock compared to the NR scenario. Based on the TEP scenario, by 2070 South Korea, the US, and the EU27 will see residential building stocks decline by 54.9%, 47.8%, and 47.4%, respectively, compared to stocks under the NR scenario. By 2070 the top two developed economies, the US and EU27, will still have an impressive 16.7 and 12.3 billion m$^2$ of floorspace, respectively, under the TEP scenario. Among emerging economies, building renovations will affect China's residential building stock strongly, lowering them by 20.4%, or 45.5 billion m$^2$, by 2070 compared to stocks under the NR scenario. The effects of building renovations on India and Africa are less pronounced, due primarily to their currently low per capita floorspace. Those two economies will continue to experience high demand for floorspace, making it difficult to compress the building stock through renovation.

In developed economies, South Korea, Canada, and Japan will be the top three in terms of lowering non-residential building stocks under the TEP scenario compared to results for the NR scenario. Declines in South Korea, Canada, and Japan will be 57.1%, 51.9%, and 41.1%, respectively, by 2070. The absolute scales of the changes, however, will be relatively small. Although US non-residential building stock in 2070 will be less based on the TEP scenario than the NR scenario, the absolute reduction will be only 4.5 billion m$^2$, resulting in a still substantial building stock of 8.2 billion m$^2$. Among emerging economies, renovations would affect China's non-residential building stock strongly, decreasing them 22.5% (5.4 billion m$^2$) more under the TEP scenario than the NR scenario. Even for the TEP scenario, non-residential buildings in India and Africa show only slight declines, because they are starting off with much less floorspace.

**Carbon intensity of global non-residential building operations**

Figure S3 shows total carbon emissions, carbon emissions per capita, and carbon emissions per floorspace of non-residential building operations from 2000 to 2021 in the 11 economies. In 2021 the total carbon emissions of the 11 economies were 2522.9 megatons of carbon dioxide (MtCO$_2$), comprising 638.8 kilograms of carbon dioxide per capita (kgCO$_2$/person) and 69.0 kilograms of carbon dioxide per square meter (kgCO$_2$/m$^2$) of floorspace. Although the US and China were the top two emitters during 2000-2021 (see Figure S3 j), China's carbon emissions represented 418.8 kgCO$_2$/person and 61.0

kgCO$_2$/m$^2$ (approximately one-seventh and one-half, respectively, of those of the US). Considering India's large population, the carbon emissions per capita were extremely low at 69.9 kgCO$_2$/person, although carbon emissions per floorspace ranked high at 93.8 kgCO$_2$/m$^2$. According to the latest information, non-residential building stocks in Australia are smaller those in other developed economies. Given this smaller amount of floorspace, Australia's carbon emissions per floorspace during 2000-2021 were significantly greater than those of other developed economies (see Figure S3 I).

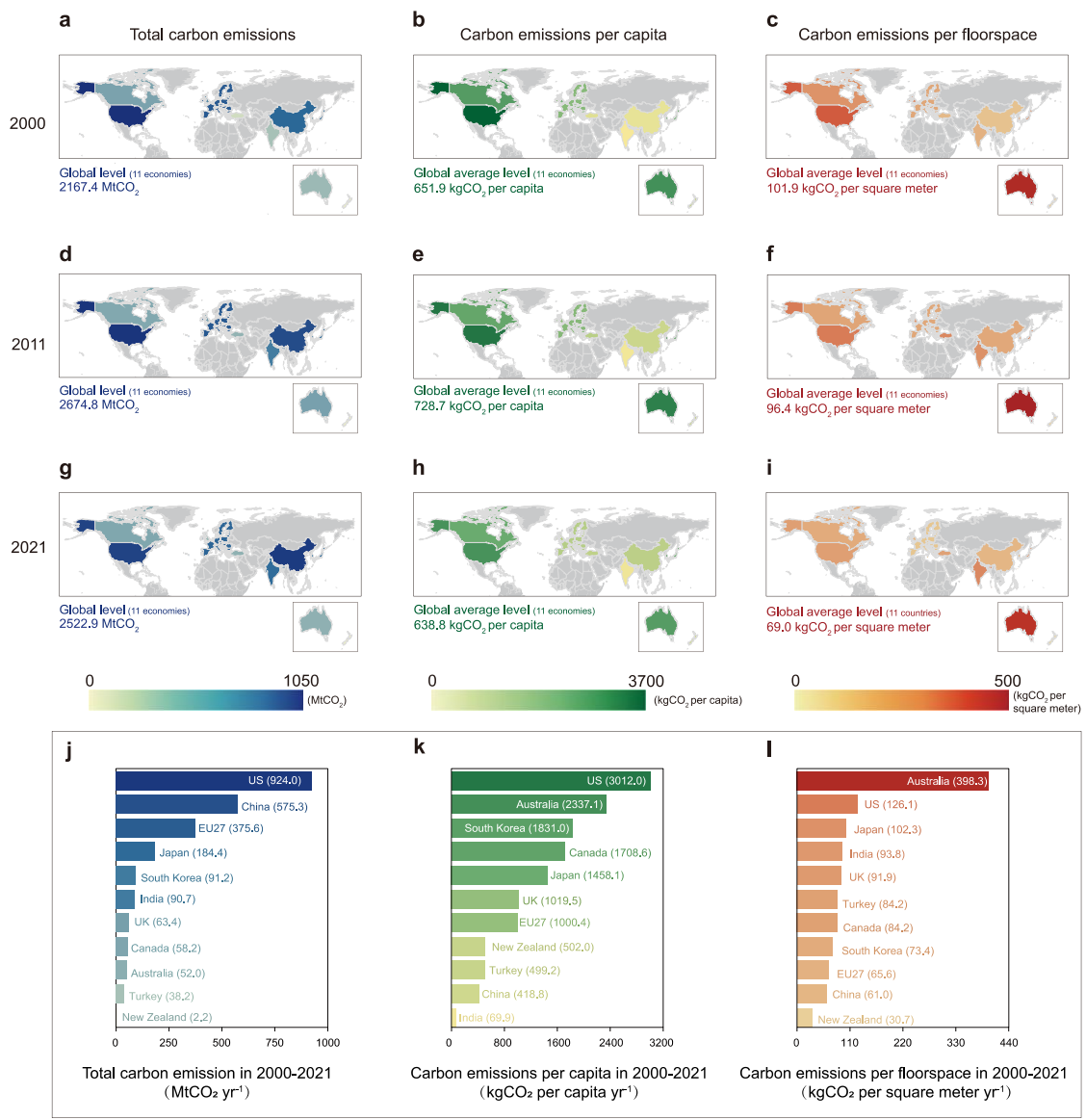

**Figure S3.** Carbon emissions from non-residential building operations in 11 economies from 2000 to 2021. Note: the carbon emission data were referred from our former study[12] and our GLOBE database (https://globe.lbl.gov/, https://globe2060.org/).

**Mitigating embodied carbon through building renovation: the case of China**

Estimating future newly contributed floorspace is important for analyzing carbon emissions associated with the building sector, especially embodied carbon. As an example, Figure S4 presents the trends in embodied carbon emissions in China's building sector (both residential and non-residential) from 2021 to 2060 based on the NR and the TEP scenarios.

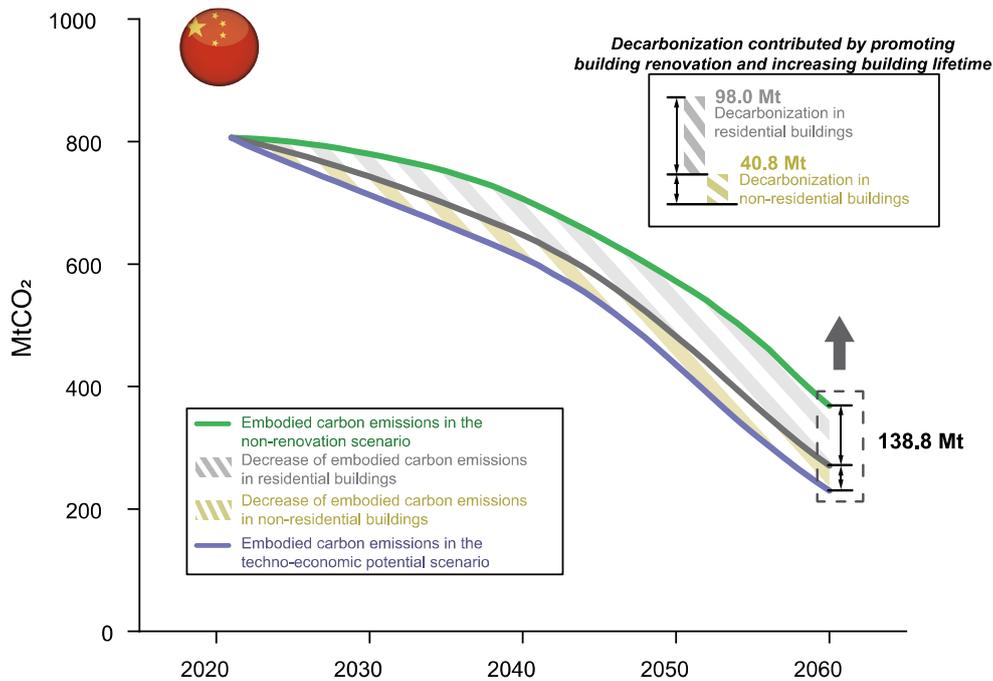

**Figure S4.** Trends in embodied carbon emissions in China's building sector from 2021 to 2060 based on the NR and TEP scenarios. Note: the embodied carbon emission data were referred from our former study.[13]

The current lifespan of buildings in China is approximately 35 years, potentially extending to 50 years for buildings constructed after 2030. Based on the NR scenario, the embodied carbon emissions of China's building sector will decrease from 806.5 to 368.5 MtCO$_2$ from 2021 to 2060. The TEP scenario, however, assumes that the renovation rate for China's residential buildings will increase at an annual rate of 1%, and for non-residential buildings will increase at an annual rate of 1.5% (starting from the current global annual average renovation rate of less than 1%). When combined with the extension of the 35-year building lifespan, from 2021 to 2060 additional renovation under the TEP scenario will lower the embodied carbon emissions of China's building sector by approximately 576.7 MtCO$_2$, which is approximately 31.7% more decarbonization than achieved in the

NR scenario. Based on the TEP scenario the extension of building lifespans through renovation will lower embodied carbon emissions substantially—by 138.8 $MtCO_2$, comprising 98.0 and 40.8 $MtCO_2$ for residential and non-residential buildings, respectively. These findings demonstrate that building renovations can play a significant role in mitigating embodied carbon emissions from the building sector, especially embodied carbon released during construction.

## 2. SUPPLEMENTAL EXPERIMENTAL PROCEDURES

The GLOBUS model (see Figure S5) operates based on assuming a specified building lifetime and rate of building renovation, and the main equations to run the GLOBUS include Eqs. 1-5 (see the main document). The scenarios and underlying assumptions are described in detail in Section 3, Scenario Setting and Assumptions. Within the GLOBUS model, building renovation refers to rehabilitating buildings that otherwise would be demolished. The model applies a renovation rate ($\alpha$) that extends buildings' lifetimes and returns them to their customary use. The formula for renovation buildings ($RB$) follows.

$$RB_{i,j,t} = \alpha \times DB_{i,j,t} \tag{Eq. S1}$$

Demolished buildings ($DB_{i,j,t}$) represent the total amount of floorspaces constructed in different years ($n$) that will be demolished in year $t$:

$$DB_{i,j,t} = \sum_{n=n_0}^{2070} DB_{i,j,t}^n \tag{Eq. S2}$$

$$DB_{i,j,t}^n = BS_{i,j,t-1}^n \times \frac{P_T^n - P_{T-1}^n}{1 - P_{T-1}^n} \tag{Eq. S3}$$

$$P_T^n = \frac{1}{\sqrt{2\pi}\sigma} \int_0^T e^{\frac{-(T-\mu)^2}{2\sigma^2}} dt \tag{Eq. S4}$$

where $T$ (or $T$-$1$) indicates the lifetime of a building constructed in year $n$ when it is demolished in year $t$ (or $t$-$1$), and $P_T^n$ shows the probability of a building being constructed in year $n$ and having a lifetime of $T$. According to current studies, the lifetime of a building follows a normal distribution based on a mean lifetime and a standard deviation of one-third of the mean building lifetime.[14,15] The demolition of renovated buildings follows a similar distribution:

$$DRB_{i,j,t}^n = RB_{i,j,t-1}^n \times \frac{P_T^n - P_{T-1}^n}{1 - P_{T-1}^n} \tag{Eq. S5}$$

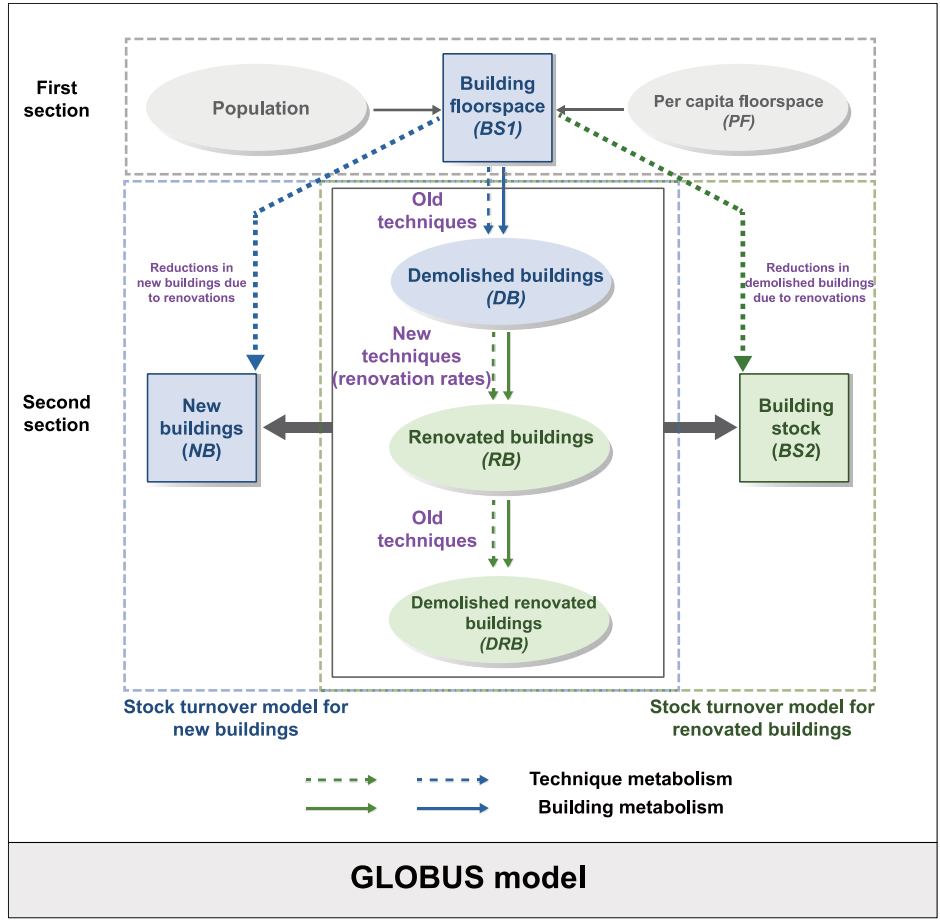

**Figure S5.** Framework of the GLOBUS model.

As for the different building stock levels in the BAU and TEP scenarios, as we mentioned previously, **the building renovation rate at the BAU level is not expected to result in a decrease in the future building stock. However, under the TEP scenario, we anticipate a decrease in the future building stock worldwide.** This projection is based on the assumption that the TEP scenario entails significant efforts towards building renovations, which may lead to a reduction in overall building stock.

Utilizing Eqs. 4 & 5 of the main document, we can prove that $BS_{i,j,t}^{BAU} - BS_{i,j,t}^{TEP} > 0$, where $BS_{i,j,t}^{BAU}$ and $BS_{i,j,t}^{TEP}$ represent the building stock under the BAU and TEP scenarios, respectively.

**Prove** $BS_{i,j,t}^{BAU} - BS_{i,j,t}^{TEP} > 0$

**Proof:**

$$BS_{i,j,t}^{BAU} - BS_{i,j,t}^{TEP} = (BS_{i,j,t}^{NR} - \sum(RB_{i,j,t}^{BAU} - DRB_{i,j,t}^{BAU})) - (BS_{i,j,t}^{NR} - \sum(RB_{i,j,t}^{TEP} - DRB_{i,j,t}^{TEP}))$$

$$= \sum((RB_{i,j,t}^{TEP} - RB_{i,j,t}^{BAU}) - (DRB_{i,j,t}^{TEP} - DRB_{i,j,t}^{BAU}))$$

**So** we just need to prove $(RB_{i,j,t}^{TEP} - RB_{i,j,t}^{BAU}) - (DRB_{i,j,t}^{TEP} - DRB_{i,j,t}^{BAU}) > 0$

**Since** the Eq. S5 in the Supplemental Information:

$$DRB_{i,j,t}^{n} = RB_{i,j,t-1}^{n} \times \frac{P_{T}^{n} - P_{T-1}^{n}}{1 - P_{T-1}^{n}} \tag{Eq. S5}$$

$\frac{P_{T}^{n} - P_{T-1}^{n}}{1 - P_{T-1}^{n}}$ represents the demolition rate of buildings built in year $n$ and demolished in year $t$ with a lifespan of $T$ (i.e., the initial lifetime or lifetime after renovation), and the value range (including the probability - $P$) is between 0 and 1.

**So** $(RB_{i,j,t}^{TEP} - RB_{i,j,t}^{BAU}) - (DRB_{i,j,t}^{TEP} - DRB_{i,j,t}^{BAU})$

$$= (RB_{i,j,t}^{TEP} - RB_{i,j,t}^{BAU}) - (RB_{i,j,t}^{TEP} \times \frac{P_T - P_{T-1}}{1 - P_{T-1}} - RB_{i,j,t}^{BAU} \times \frac{P_T - P_{T-1}}{1 - P_{T-1}})$$

$$= (RB_{i,j,t}^{TEP} - RB_{i,j,t}^{BAU}) \times (1 - \frac{P_T - P_{T-1}}{1 - P_{T-1}})$$

**Since** the Eq. S1 in the Supplemental Information:

$$RB_{i,j,t} = \alpha \times DB_{i,j,t} \tag{Eq. S1}$$

And the building renovation rates $\alpha^{TEP} > \alpha^{BAU}$

**So** $(RB_{i,j,t}^{TEP} - RB_{i,j,t}^{BAU}) - (DRB_{i,j,t}^{TEP} - DRB_{i,j,t}^{BAU})$

$$= (RB_{i,j,t}^{TEP} - RB_{i,j,t}^{BAU}) \times (1 - \frac{P_T - P_{T-1}}{1 - P_{T-1}})$$

$$= (\alpha^{TEP} \times DB_{i,j,t} - \alpha^{BAU} \times DB_{i,j,t}) \times (1 - \frac{P_T - P_{T-1}}{1 - P_{T-1}})$$

$$= DB_{i,j,t} \times (\alpha^{TEP} - \alpha^{BAU}) \times \left(1 - \frac{P_T - P_{T-1}}{1 - P_{T-1}}\right)$$

**As** $DB_{i,j,t} > 0$, $\alpha^{TEP} - \alpha^{BAU} > 0$, and $1 - \frac{P_T - P_{T-1}}{1 - P_{T-1}} > 0$

**So** $(RB_{i,j,t}^{TEP} - RB_{i,j,t}^{BAU}) - (DRB_{i,j,t}^{TEP} - DRB_{i,j,t}^{BAU}) > 0$

**In summary,** $BS_{i,j,t}^{BAU} - BS_{i,j,t}^{TEP} > 0$

Due to $BS_{i,j,t}^{BAU} - BS_{i,j,t}^{TEP} > 0$, considering the building stock under the BAU scenario

keeps unchange, **the building stock under the TEP scenario will decrease in the future.**

Overall, **we believe the global building stock to generally increase under the NR scenario, remain unchanged under the BAU scenario, and decrease under the TEP scenario,** as illustrated by Figure 3 of the main document.

## 3. SCENARIO SETTING AND ASSUMPTIONS

In this study, the **non-renovation (NR)** scenario estimates the stocks of residential and non-residential buildings in various economies based on the population size and per capita floorspace (see Eq. 1 of the main document) without considering building renovations.[16,17] (That is, the building renovation rate is 0.) The building stock in the NR scenario was established based on historical and forecast data and research findings.[18,19] The historical and predicted population data for many economies were sourced from https://www.populationpyramid.net/; predicted populations for the EU27 were obtained from Eurostat. The historical per capita floorspace data were derived from the *Energy Efficiency Indicators* released by the IEA. Other sources are described below.

1. **China:** To assess per capita floorspace in China, we referred to the *Study on the Implementation Path of Carbon Peak and Carbon Neutrality in the Building Sector* released by China's Ministry of Housing and Urban-Rural Development in 2021. That study led us to set China's per capita floorspace in 2070 at 52.6 m$^2$ for residential buildings and 22 m$^2$ for non-residential buildings.[20]

2. **US:** The U.S. Energy Information Administration's *Annual Energy Outlook 2023* predicts that in 2050 the number of households in the US is projected to reach 157.7 million, with an average house size of 1916.3 square feet. Besides, the non-residential building stock will be 123.3 billion square feet.[4] In other words, by 2050 the per capita floorspace of residential buildings in the US will reach 74.8 m$^2$, and the per capita floorspace of non-residential buildings will reach 30.5 m$^2$.

3. **Canada:** The IEA forecasted that the residential building stock in Canada will be 950 million m$^2$ greater in 2050 than in 2018, as described in the *Heating and Cooling Strategies in the Clean Energy Transition*.[1] By 2050 the per capita floorspace of Canadian residential buildings will reach 67.8 m$^2$, and that of non-residential buildings will reach 21.3 m$^2$.

4. **The United Kingdom (UK):** Several native British researchers concluded that residential floorspace in the UK will reach 2906 million m$^2$ by 2050.[8] They estimated that by 2050 per capita floorspace will reach 40.5 m$^2$ in residential buildings and 10.0 m² in non-residential buildings.

5. **EU27:** The EU's SENTINEL Project, which promotes the low-carbon transition of the EU energy system, estimated that the EU's floorspace will increase 16% by 2050.[21] Specifically, by 2050 the per capita floorspace of residential buildings and non-residential buildings in the EU27 will reach 56.9 m² and 20.3 m², respectively.

6. **Japan:** Japanese scholars believed that despite the decrease in the country's average household size, the per capita floorspace of Japan's residential buildings will reach approximately 49 m² by 2050, an increase of 21% compared to 2013. Besides, the year 2050 will see 19.5 m² per capita for non-residential buildings.[7]

7. **South Korea:** Researchers from South Korea's Green Energy Strategy Institute believed that by 2050 the per capita floorspace of residential buildings in South Korea will increase to approximately 41.9 m², and the per capita floorspace of non-residential buildings approximately 39.2 m².[9]

8. **India:** Indian scholars at Lawrence Berkeley National Laboratory assessed that by 2050 the floorspace of urban residential buildings in India will reach 21.5 billion m² and that of non-residential buildings 3.6 billion m².[22] In other words, by 2050 the per capita floorspace of residential buildings in India will increase to 28.9 m²; per capita floorspace for non-residential buildings will reach 2.2 m².

9. **Africa:** The IEA's report titled *Perspectives for the Clean Energy Transition* indicated that building floorspace in Africa will reach 58 billion m² by 2050;[2] the per capita floorspace for residential buildings will increase to 21.0 m² rapidly and that of non-residential buildings 2.3 m².

10. **LAC:** The Global Alliance for Buildings and Construction's 2020 *Regional Roadmap for Buildings and Construction in Latin America*, published by the IEA, stated that by 2050 floorspace in LAC is expected to have grown by 65%, dominated by almost 11 billion m² in residential buildings.[3] Specifically, by 2050 the per capita floorspace of residential buildings and non-residential buildings in LAC will reach 35.7 and 6.8 m², respectively.

11. **Australia:** Based on their evaluation, Australian native scholars expected that by 2050 residential buildings will reach a stock of 2.6 billion m² and non-residential buildings a stock of 221.7 million m$^2$.$^{23,24}$ Furthermore, the per capita floorspace in residential buildings and non-residential buildings will reach 80.8 and 6.9 m$^2$, respectively, by 2050.

12. **New Zealand:** According to the 2016 *Global Status Report* by the United Nations Environment Programme, the annual growth rate of building floorspace in New Zealand will be 2% from 2016 to 2030, declining to 1% from 2031 to 2050.$^5$ By 2050 the per capita floorspace of residential buildings will reach 66.0 m$^2$, and of non-residential buildings 20.6 m$^2$.

13. **Turkey:** Native Turkish scholars projected that Turkey's total building floorspace will reach 5.86 billion m$^2$ by 2050, with residential building floorspace reaching 43.7 m$^2$ per capita.$^{25}$ They estimated per capita floorspace of non-residential buildings in Turkey to be 12.3 m$^2$ by 2050.

14. **Indonesia:** A report titled the *Roadmap for an Energy Efficient, Low-carbon Building and Construction Section in Indonesia*, issued by the Global Alliance for Buildings and Construction, estimated that Indonesia's building floorspace will increase 4% annually until 2030, after which it will keep approximately 3% to 5% annually until 2050.$^{26}$ The per capita floorspace of residential buildings was projected to reach 16.7 m$^2$, and that of non-residential buildings 6.9 m$^2$, by 2050.

**The business-as-usual (BAU)** scenario constitutes one of two additional scenarios incorporated into the building turnover model (GLOBUS). Building on the NR scenario, the BAU scenario assesses the current level of building renovation in each economy. The model assumes that the current low rates of building renovation will persist. Because building renovation in developed economies already has reached a substantial scale, the BAU scenario predicts a higher volume of renovation than forecasted by the NR scenario (see Eqs. 2 & 4 of the main document).

In developing economies, building renovations currently command a minimal share of new materials. In practical terms, the NR and BAU scenarios examined in this study reveal identical renovation rates for developing economies, specifically China, India, and Africa. Table S1 lists the parameter settings for the BAU scenario.

**Table S1.** Parameter settings of the GLOBUS model for the BAU scenario.

| Economy | Average lifetime (years) | Initial lifetime (years) | Renovation cycle (years) | First renovation | Rate of first renovation | Second renovation | Rate of second renovation |
|---|---|---|---|---|---|---|---|
| US | 75 [27] | 25 | 25 | √ | Residential: 10%-49.2% <br> Non-residential: 10%-34.5% [28] | √ | Residential: 10%-19.8% <br> Non-residential: 10%-14.9% [28] |
| Canada | 75 [27] | 25 | 25 | √ | Residential: 0.6%-34.9% <br> Non-residential: 0.6%-69.2% [29] | √ | Residential: 0.6%-5.5% <br> Non-residential: 0.6%-10.4% [29] |
| EU27 | 120 [30] | 40 | 40 | √ | Residential: 12.0%-61.0% <br> Non-residential: 9.0%-33.5% [31] | √ | Residential: 12.0%-21.8% <br> Non-residential: 9.0%-13.9% [32] |
| Japan | 50 [33] | 25 | 25 | √ | Residential: 1.4%-35.7% <br> Non-residential: 1.4%-35.7% [34] | | |
| South Korea | 50 [35] | 25 | 25 | √ | Residential: 3%-37.3% | | |

| Economy | Average lifetime (years) | Initial lifetime (years) | Renovation cycle (years) | First renovation | Rate of first renovation | Second renovation | Rate of second renovation |
|---|---|---|---|---|---|---|---|
| | | | | | Non-residential: 3%-37.3% [36] | | |
| China | Before 2030: 35 [37] <br> 2031-2070: 50 [38] | Before 2030: 35 <br> 2031-2070: 50 | | | | | |
| India | Before 2030: 35 <br> 2031-2070: 50 [39] | Before 2030: 35 <br> 2031-2070: 50 | | | | | |
| Africa | Before 2030: 35 <br> 2031-2070: 50 | Before 2030: 35 <br> 2031-2070: 50 | | | | | |

The **techno-economic potential (TEP)** scenario assumes a substantial increase in building renovation rates propelled by continuous technological and economic development (see Eqs. 3 & 5 of the main document), compared with the renovation rates in the BAU scenario. Cultural differences related to buildings result in European and American buildings generally having longer lifetimes than those in Asia. The analysis undertaken in this study assumes that buildings in the US, Canada, and the EU27 will undergo demolition after two renovation cycles. The detailed assumptions central to the

TEP scenario are outlined in Table S2.

Table S2. Parameter settings in the GLOBUS model for the TEP scenario.

| Economy | Average lifespan (years) | Initial lifetime (years) | Renovation cycle (years) | First renovation | Rate of first renovation | Second renovation | Rate of second renovation |
|---|---|---|---|---|---|---|---|
| US | 75 [27] | 25 | 25 | √ | Residential: 10.0%-88.4% Non-residential: 10.0%-59.0% [40,41] | √ | Residential:10.0%-29.6% Non-residential: 10.0%-19.8% [40] |
| Canada | 75 [27] | 25 | 50 | √ | Residential: 0.6%-69.2% Non-residential: 0.6%-98.6% [29] | √ | Residential: 0.6%-10.4% Non-residential: 0.6%-20.2% [29] |
| EU27 | 120 [30] | 40 | 40 | √ | Residential: 12.0%-100.0% Non-residential: 9.0%-58.0% [31,42] | √ | Residential: 12.0%-31.6% Non-residential: 9.0%-18.8% [32,42] |
| Japan | 50 [33] | 25 | 25 | √ | Residential: 1.4%-70.0% Non-residential: 1.4%-70.0% [34] | | |

| Economy | Average lifespan (years) | Initial lifetime (years) | Renovation cycle (years) | First renovation | Rate of first renovation | Second renovation | Rate of second renovation |
|---|---|---|---|---|---|---|---|
| South Korea | 50 [35] | 25 | 25 | √ | Residential: 3%-100% Non-residential: 3%-100% [36,43] | | |
| China | Before 2030: 35 2031-2070: 80 | Before 2030: 35 [37] 2031-2070: 50 [38] | 30 | √ | Residential: 0-49% Non-residential: 0-73.5% [44] | | |
| India | Before 2030: 35 2031-2070: 80 | Before 2030: 35 2031-2070: 50 [39] | 30 | √ | Residential: 0-49% Non-residential: 0-73.5% | | |
| Africa | Before 2030: 35 2031-2070: 80 | Before 2030: 35 2031-2070: 50 | 30 | √ | Residential: 0-49% Non-residential: 0-73.5% | | |

Based on the aforementioned assumptions, we use the NR scenario as a case study. The validation and uncertainty analysis of the data on global floorspace per capita for residential and non-residential buildings from 2000 to 2070 are illustrated as follows.

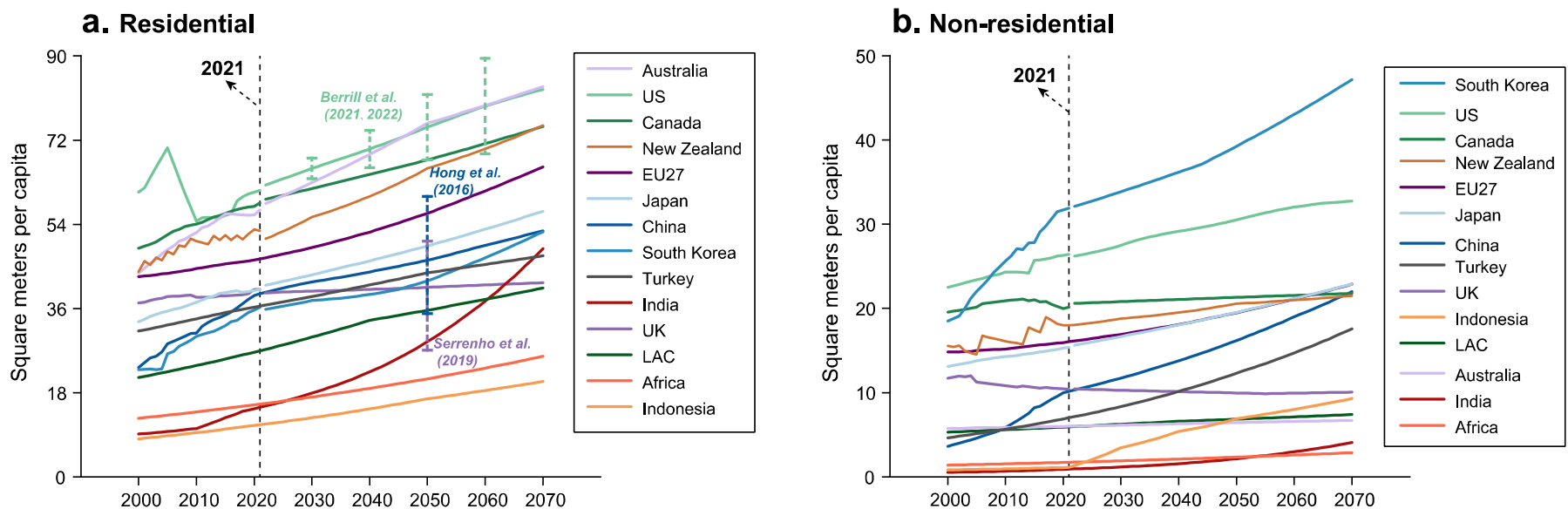

**Figure S6.** Validation and uncertainty analysis of the data on global floorspace per capita for (a) residential and (b) non-residential buildings from 2000 to 2070 based on the NR scenario. Note: the literature referenced in this graph include sources[14, 45-47].

# REFERENCES FOR SUPPLEMENTAL INFORMATION